# Micro-arcsecond Astrometry Technology: Detector and Field Distortion Calibration

Michael Shao[1], Chengxing Zhai[1], Bijan Nemati[2], Inseob Hahn[1], Russell Trahan[1], and Slava Turyshev[1]

1. Jet Propulsion Laboratory, California Institute of Technology, 4800 Oak Grove Dr, Pasadena, CA 91109
2. Tellus1 Scientific, LLC, Madison, AL 35758

Abstract

Microarcsecond (uas) astrometry is an indispensable technique to detect earth-like exoplanets, fully characterize exoplanetary orbits, and measure their masses – information critical for assessing their habitability. Highly accurate astrometric measurements can also probe the nature of dark matter, the early universe, black holes, and neutron stars, thus providing unique data for new astrophysics. This paper presents technologies of calibrating detectors and field distortions for achieving narrow field uas relative astrometry with a focal plane array detector on a 6 m telescope.

## 1 Introduction

Gaia mission has revolutionized astrophysics by providing extremely accurate global reference astrometry. Going beyond Gaia to achieve narrow field microarcsecond (uas) astrometry enables to detect earth-like exo-planets by measuring the reflex motion of the host stars (Unwin et al. 2008). Even though popular methods like radial velocity (RV) and transit have successfully discovered thousands of exo-planets, only the astrometric detection method would allow us to fully determine the orbits and measure the masses of the exo-planets in general[1]. The mass of an exoplanet is a crucial parameter for determining whether the planet is suitable for hosting life because its atmosphere and geophysical processes strongly depend on the mass. Compared with the RV method, astrometric detection is less affected by perturbations due to stellar activities and has better sensitivity for longer period exo-planets, thus complementary to the RV and transit methods. For this unique role, NASA has listed "Stellar Reflex Motion Sensitivity – Astrometry" as a Tier 1 Technology Gap (NASA Strategic Technology Gaps) for measuring the masses of habitable exoplanet targets.

Besides exo-planet sciences, uas astrometry can be used to study the nature and distribution of dark matter by accurately measuring stellar proper motions. In addition, highly accurate astrometric measurements could allow us to study black holes and neutron stars in the investigations of black holes mergers, study of X-ray binaries, and detection of microlensing effects. Measuring the microlensing effects of primordial black holes and coherent proper motions at the large scale can also be used to uncover key information about the early Universe.

---

[1] RV can only measure the product of mass and the sine of the orbit inclination. A face-on orbit would have zero RV signals. Direct imaging might help determine the orbit inclination, but the inner working angle may limit this approach to only a small subset of the geometric configurations.

Gaia's end-of-mission accuracy is 10-20 uas (Lindegren L. et al. 2020a, 2020b). The best Hubble Space Telescope accuracy is 20-40 uas (Riess et al. 2014). The Space Interferometry Mission (SIM) in the 2000s was the first mission attempting to perform uas astrometry using stellar interferometry. For SIM, sophisticated calibration had to be developed to correct for systematic errors such as stellar color effects needed to achieve measurements at the uas-level of accuracy (Uwin et al. 2008; Milman et al. 2007; Zhai et al. 2007; Zhai 2009a; Zhai et al. 2009b). Unfortunately, SIM did not go on as a flight project, thus to date there are no uas astrometric capabilities available for astrophysics.

Modern focal plane array CCD and CMOS detectors offer accurate measurements of photon fluxes with very low read noise over a regular array, typically much larger than 1K x 1K, with the pixel as small as a few microns. Working with a diffraction-limited large space telescope is a natural choice for the next generation of accurate space astrometry. The Near-Earth Astrometric Telescope (NEAT) (Malbet et al. 2012) and the recent Theia (Malbet et al. 2022) are two mission concepts proposed to the European Space Agency to perform uas-level astrometry using a meter-class telescope with a focal plane array detector. NASA's priority flagship mission for the next decade will be a 6 m telescope for observing habitable exoplanets and in the search for new physics. For exo-planet sciences, ability to determine masses of habitable exoplanetary targets is crucial. All these mission concepts call for technologies to calibrate detectors and optical field distortions to achieve uas accuracy needed for reduction of the systematic errors due to imperfect detectors and optics down to sub-uas. A search for Earth-like planets in the habitable zone of nearby FGK stars means looking at bright nearby stars. Many of these stars will saturate the detector on a 6 m space telescope. When a bright star is saturated, the diffraction spikes can be used to locate the star for astrometry. We therefore must consider doing astrometry using the diffraction spikes of these saturated stars in the image.

This article presents our research work on calibration technology to characterize pixel responses of array detectors and field distortions for achieving uas astrometry. We also discuss future works needed to demonstrate the viability of this technology. In Section 2, we describe our calibration architecture. We present results in Section 3. In Section 4 we conclude and discuss directions for future work.

## 2    Calibration Architecture

If observed from 10 pc, the reflex motion of the Sun under the gravitational pull from the Earth results in an astrometric displacement with magnitude of 0.3 uas. Assuming an observing cadence of 10 measurements over 5 years, the end-of-mission accuracy needed to detect such a planetary system at an SNR of 6 is 0.05 uas. The single measurement accuracy needed is sqrt(50)x0.05uas ~ 0.35uas. Astrometric errors consist of both random and systematic errors. Random errors mainly come from photon shot noise, detector read noise, dark currents, and zodi background noise and can be mitigated by collecting enough signal. Systematic errors, on the other hand, require accurate and viable calibration methods to correct.

Two major sources of systematic errors for accurate astrometry are those due to imperfect detectors with non-uniform pixel geometry and optical distortions due to optics with aberrations and misalignments. To calibrate for the pixel geometry variations and to account for semiconductor fabrication errors, we use laser metrology and apply a low-order polynomial model to calibrate field distortions by observing a dense star field with systematic dithers of the field of view (FOV). Because target stars are typically nearby stars, they are bright and may saturate the detector for a 6 m telescope. We shall also handle the special case of centroiding saturated stars by locating their diffraction spikes.

Several groups have tried similar approach in the past using HST images and more importantly Gaia images, where absolute astrometric accuracy of ~100 uas was claimed. The Gaia detector saturates for stars brighter than G ~6 mag (Sahlmann et al. 2016). It should be mentioned that Gaia is an absolute astrometric instrument while exoplanet astrometry is relative astrometry. That is if centroiding the star using the Airy disk has a slight offset from centroiding using the diffraction spike, that offset is not a problem for exoplanet detection if it does not vary between epochs as we are only interested in the motion of the star not its absolute position in the sky. However, this offset may not be stable and could depend on the field, therefore it is important to calibrate this offset. Our calibration technique will be described in the context of the system architecture presented in the next subsection.

## 2.1  System architecture

Our study is based on the Theia mission design, which is a single spacecraft carrying a meter-class or larger Korsch three-mirror anastigmatic (TMA) telescope [Malbet et al. 2022]. This can be scaled to a larger telescope of 6 m. To have a sufficiently large FOV, it is necessary to have tertiary optics to correct for wavefront aberrations needed to achieve good imaging quality. An astrometric telescope produces images of the sky and therefore maps the sky into the detector's pixel coordinate. An ideal system could be modeled geometrically by mapping the celestial sphere coordinate, right ascension (RA), and declination (Dec), into an imaging plane sampled by a uniform rectangular pixel coordinate in terms of rows and columns. A realistic system contains both random and systematic astrometric measurement errors. For a 6 m telescope considering here, an integration time of 10 min provides sufficient number of photons that can be used to average down random errors to a sub-uas level for targets brighter than a $12^{th}$ mag star. It is crucial to calibrate the systematic errors from the imperfection of the detector and optics, which we shall describe in detail.

## 2.2  Calibration methods

The dominant imperfection is pixel geometry errors. For this, we have developed laser metrology that projects laser fringes on the focal plane array (which might be a mosaic of chips) to measure the responses in the Fourier domain. A study based on simulation has shown this method can perform centroiding at the micro pixel level (Zhai et al. 2011), which is sufficient for a sub-uas astrometric accuracy. This is possible because the pixel scale must be less than 0.1 arcsec to critically sample diffraction limited point-spread-functions (PSFs) from a meter-class optical telescope. Imperfect optics comes from aberrations due to field distortions, wavefront errors due to misalignment, and field-dependent footprint on the optics that is present mainly due to beam walk on the tertiary optics.

### 2.2.1 Detector calibration using focal plane metrology

Normal CCD/CMOS detector calibration measures the QE, dark current, and read noise of each pixel and assumes the pixels are perfectly spaced in a rectangular grid. QE gradients within a pixel and geometric errors in the placement of pixels will result in centroiding errors of ~1e-3 pixels. We describe below a calibration procedure that measures the position of every pixel relative to a regular grid with better than 1e-4-pixel accuracy. For a 4um pixel, this is equivalent to measuring the X and Y positions of the pixel to 0.4 nm. The calibration technique uses laser light launched from the tips of fibers to illuminate the focal plane (Figure 1). Two fibers' illumination creates fringes on the pixels. If the fiber ends are attached to a thermally stable block, the fringe spacing is then a stable reference for the metrology of the pixels. The fringes can be made to move across the focal plane by shifting the phase of the light launched from one of the fibers relatively to the other. The intensity variations detected by a pixel can be used to determine the fringe phase at the pixel, thus the effective location of the pixel. With the moving fringes we can calibrate the pixel geometry of all the pixels at the same time.

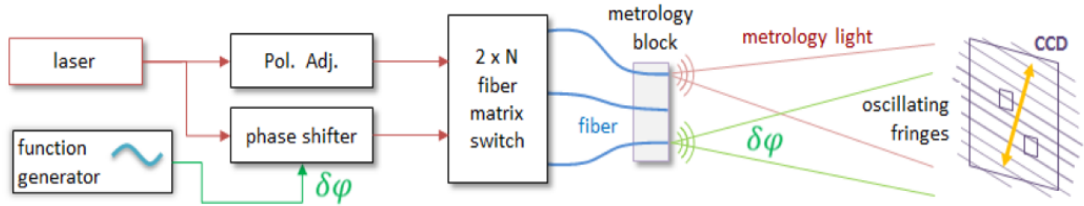

Figure 1: High-precision calibration of focal-plane errors uses moving fringes placed on the detector. Each pixel's location can be derived from the measured phase and amplitude of the fringe at that pixel.

### 2.2.2 field distortion calibration

The key to a successful field distortion calibration is a distortion model that can describe the field distortion with sub-uas errors. Theoretically, we know that the field distortions from a perfect optical system can be modeled as lower-order polynomials. Based on simulations, we found that optical field distortions from optical systems with wavefront aberrations can all be modeled in terms of low order polynomials to sub-uas (as shown in section 3.2). Such aberrations include those due to non-ideal optics with peak-to-valley wavefront errors of $\lambda/20$, misalignment errors at the level of 1 arcsecond, and those due to the beam walks on tertiary optics. This is well understood as the wave propagations near the optical axis tend to smooth out the effect of wavefront aberrations of higher spatial frequency on the optics.

## 2.3 Error budget for uas astrometry

Table 1 shows the single measurement error budget for 1 m and 6 m telescopes respectively, with integration times of 1 hour and 10 minutes, which are determined by the requirement for averaging down random astrometric errors.

Table 1. Error Budget for uas Astrometry

| Telescope diameter | Integration Time (sec), | random error (uas) | Detector Calibration (uas) | Optical Distortion (uas) | Total (uas) |
|---|---|---|---|---|---|
| 1 m | 3600 | 0.44 | 0.3 | 0.3 | 0.61 |

| 6 m | 600 | 0.13 | 0.066 | 0.066 | 0.18 |

## 3 Results

In this section, we present our results on the calibration techniques based on lab experiments and simulations as a preliminary validation of our methodology.

### 3.1 Detector Calibration

#### 3.1.1 Pixel geometry calibration

Pixel geometry calibration can be performed using the laser metrology shown in Fig. 1. The leading order inter-pixel response variations are pixel QE (flat field response) and effective pixel locations in the array, which can deviate from a regular grid. We have characterized an E2V

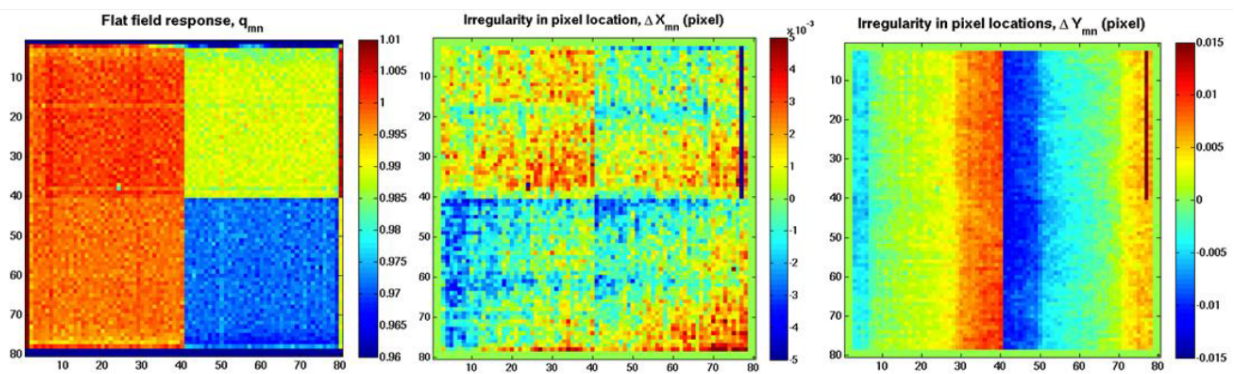

Figure 2: (Left) Flat-field response; (Center) Pixel-location irregularity (row, X); (Right) pixel location irregularity (column, Y).

CCD39 with an array size of 80x80 for flat-field response and x– and y–direction pixel-location irregularity. The left plot in Fig. 2 displays relative QEs of 80x80 pixels. The mid and right plots display the pixel irregularity as deviations in X (row) and Y (column). This particular 4-quadrant sensor showed a very obvious "step and repeat" error of a few percent of a pixel in the column direction. The pixel location measurements reach a precision of 1e-4 pixels with an integration time of about 100 seconds (Nemati et al. 2011). This particular CCD has 24um pixels with about 700nm pixel placement error between the left and right half of the chip and random pixel-to-pixel location errors on the 40-50nm scale.

#### 3.1.2 Accuracy of differential centroiding of pseudo-stars

Astrometry is the measurement of the angular distance between stars, and the brightness centroid is an effective measure of the position of a stellar image. Inter-pixel response variations directly affect centroiding stars in the field. The ultimate test of the accuracy of focal-plane calibration and centroiding is an astrometric validation experiment, which we now describe. The focal plane is illuminated with some pseudo stars, in this case, generated by imaging a fiber bundle onto the sensor shown in Fig. 3, where three pseudo-star images appear on the camera which has been calibrated. The camera is then moved while frames are being taken so that the images fall on different regions of pixels. Assuming that the pseudo-star images are stable relative to each other, the measured separations should be independent of the detector position.

For the micro-pixel level of precision, it is necessary to take care of effects arising from pixelated images and the non-ideal response of the focal-plane array pixels. Regarding the pixelated images, we are aided in this by the fact that the stellar images are bandwidth-limited due to the finite aperture of the telescope because telescope images have no structure finer

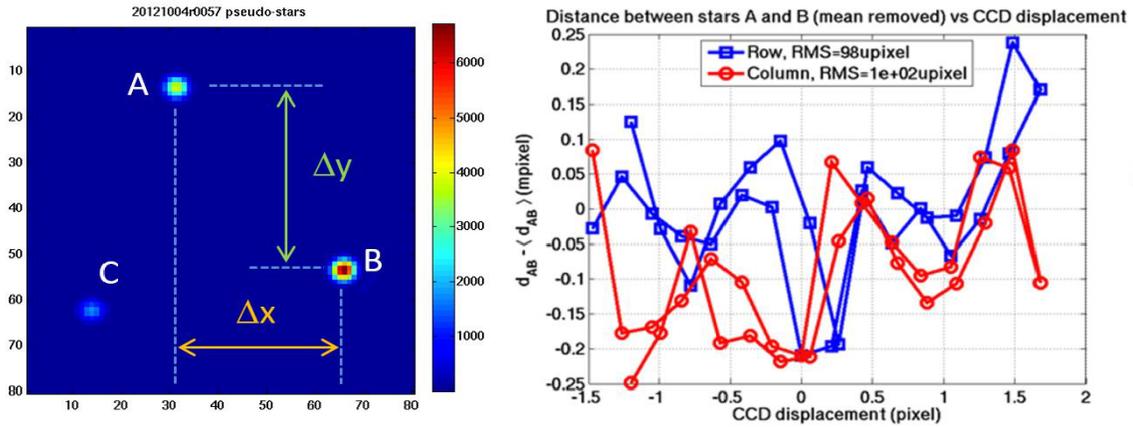

Figure 3: (Left) An astrometric test measures the consistency of inter-star distances on the focal plane as the line of sight is changed. (Right) Results of an astrometric test: centroid distance between pseudo-stars A and B in row and column directions, with mean removed, versus the displacement of the CCD

than ∼ λ/D (projected onto the sky). If we critically sample the focal plane (*i.e.,* have > 2 pixels per λ/D projected onto the sky), the PSF shape can be described without loss by its Fourier transform. We use the Fourier technique to shift images in a lossless manner, and cross-correlate them to accurately estimate the distance between the stars.

The right plot in Fig. 3 shows the experimental results of the variations of the measured distance between stars A and B when we displaced a calibrated CCD. Since the separation of the pseudo-stars A and B is stable, the variations in the measured distance between A and B are due to errors. Without any calibration, this error can be as large as 1 milli-pixel. The calibration reduced the error to about 100 μpix per step. Averaging 10 steps can further reduce the error roughly by a factor of 1/sqrt(10). Our operational concept calls for dithering the system pointing to take advantage of the averaging.

## 3.2   Field distortion Calibration

Field distortion in a telescope means that stars imaged by the telescope do not appear in the locations corresponding to the angular positions of the stars with perfect fidelity as a geometric projection from the plane of sky. A telescope with optical elements having zero wavefront error can still have significant distortion. In our telescope design, which is symmetric, the distortion is only in the radial direction, see Figure 13 in (Nemati et al. 2020). We conducted simulations tracing millions of rays at ∼10,000 points in the FOV and found that the radial distortion can be modeled to very high accuracy, < 1μas, with a ninth-order polynomial as shown in Fig. 4.

### 3.2.1 Field distortion from misalignment

If we perturb the optical alignment by a small amount, the image would be, in general, still diffraction-limited. The distortion map, however, would be no longer circularly symmetric and had both radial and azimuthal terms. We found that the distortions for an imaging system with optics with zero wavefront errors and 1-arcsecond alignment error can be fitted with a low-order 2D polynomial with centroiding errors less than 1e-5 pixel (Malbet et al. 2022).

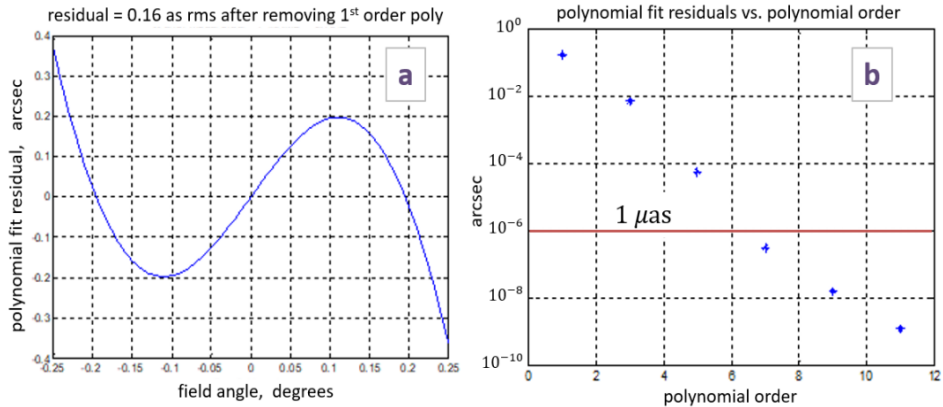

Figure 4: Modelling optical distortion and its removal: a) the residual radial field distortion after removing a linear radial trend; b) the rms error, in arcseconds, after removing a polynomial fit of successively higher orders. The line at 1 µas represents a reference point and also the approximate allocation to this error.

### 3.2.2 Field distortion from optics fabrication errors.

Diffraction-limited optics are manufactured to very tight tolerances. But when centroiding a star's position to 1e-5 of the diffraction limit, even $\lambda/20$ p-v (peak-to-valley) wavefront errors can result in significant biases. Wavefront errors on the primary would produce changes in the PSF, but do so in the same way for all stars in the FOV, thus would not introduce distortions. The wavefront errors on subsequent surfaces, however, would be sampled differently by stars in different parts of the FOV, which would lead to field-dependent centroiding errors, thus distortions. Optical surface wavefront errors at or near the image plane would not result in significant optical distortions. For a three-mirror anastigmat (TMA) telescope (Nemati et al. 2020, Malbet et al. 2022), this beam walk is the largest on the curved tertiary mirror. In this section, we evaluate the astrometric distortions caused by a tertiary mirror that has a $\lambda/20$ p-v wavefront error, where the p-p (peak-to-peak) beam walk is 50% of the diameter beam (see Fig. 5).

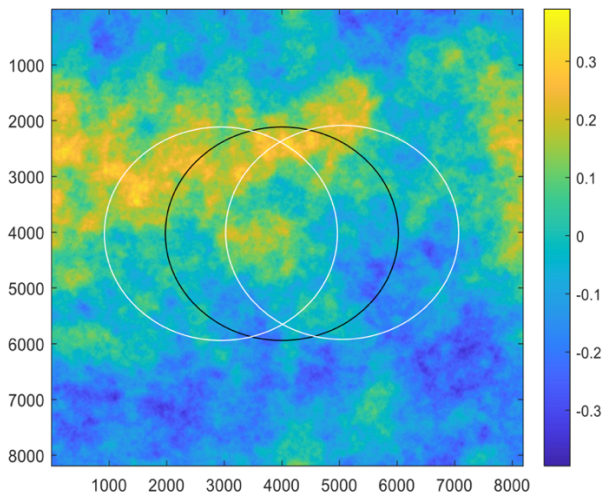

Figure 5, Generated phase errors (rad) as an 8K x 8K array representing the optical surface of the tertiary mirror, where the two circles of diameters of 4K points represents two optical footprints on the mirror surface at the opposite sides of the FOV.

High-quality optics typically have wavefront errors with p-v amplitude of $\lambda/15$ p-v

(about +/- 0.2 radians in phase) with a $1/f^3$ spatial frequency power spectrum (or $1/f^{1.5}$ amplitude spectrum). We simulate a set of wavefront errors by first generating Gaussian white noises sampled at 8K x 8K grid points and then applying a low pass filter with $1/f^{1.5}$ response and scaling the p-v value to $\lambda/15$. We found that the RMS of wavefront errors is about $\lambda/100$. While a wavefront error of $\lambda/15$ p-v (~40nm p-v for $\lambda$ = 633nm) represents a high-quality diffraction-limited optic (the state-of-the-art optics for EUV lithography is a diffraction-limited optic at $\lambda$=13nm.).

We sample the beam footprint as a circle with a diameter of 4000 points, *i.e.*, each star over the FOV samples its own 4000-point diameter circle on the tertiary mirror as illustrated in Fig. 5. For each position in the FOV, we fit a plane to the phase error in the 4000-point circle of the corresponding footprint. The tilt of the plane represents the leading order centroid shift due to this wavefront error. Figure 6 plots the centroid shift in the X and Y directions over the FOV.

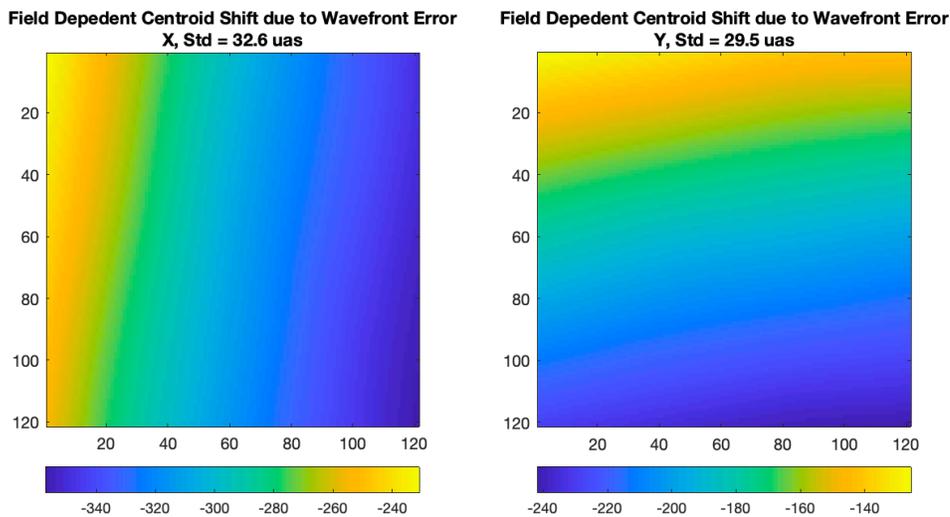

Figure 6. Field-dependent centroiding error due to beam walk on the tertiary optics with wavefront error of lambda/15 (p-v).

The centroid shift is a slowly varying distortion function over the field with a p-v range of 100 uas. It can be modeled by a two-dimensional polynomial model and calibrated by using reference stars in the FOV. A relatively low-order polynomial can model this type of distortion because the wavefront error is averaged over the beam, and the beam walk is limited to 50% of the beam diameter p-v. **Error! Reference source not found.** shows the distortion error residual RMSs after fitting two 2D polynomials as functions of the order of the 2D polynomials.

Figure 8 shows centroiding residuals over the field after fitting a 15$^{th}$ order 2D polynomials. We note that the residuals are within +/-0.1uas.

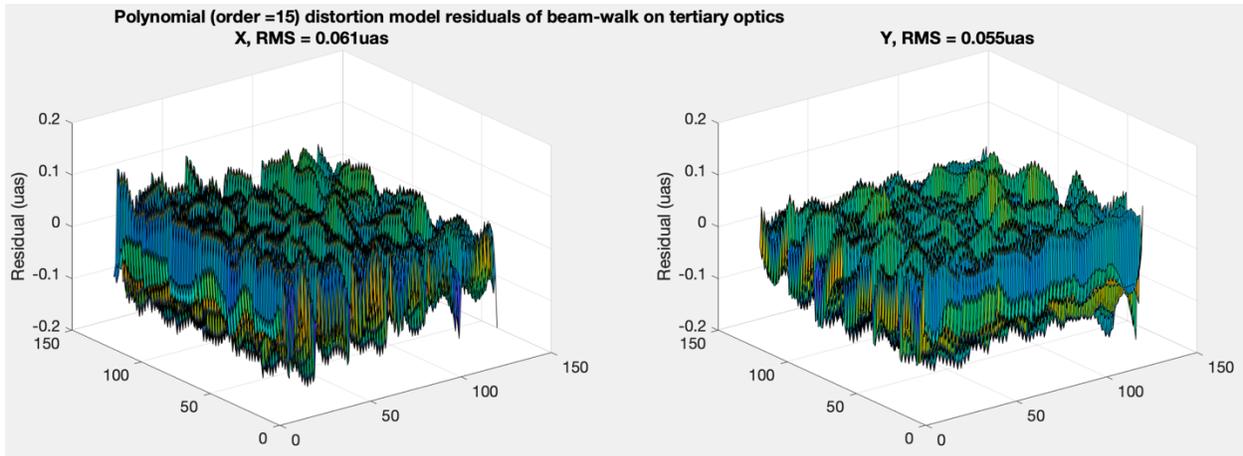

Figure 8. Distortion Residuals over the field using a 15th order polynomial model.

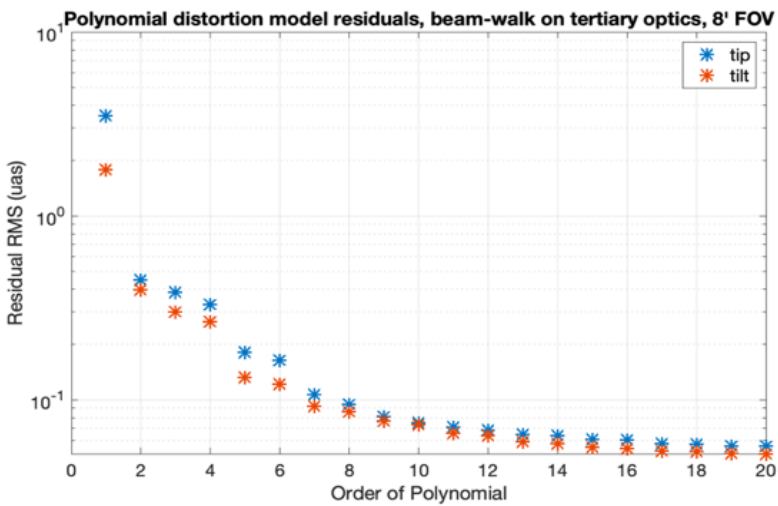

Figure 7. RMS of field distortion residuals as function of the order of the polynomial model.

### 3.3 Centroiding using diffraction spikes.

In astrometric detection of exoplanets, quite often, the target star is very bright (~0-8 mag) while the reference stars are dim (~12-19 mag). For the brightest nearby stars, the image will saturate the detector. Fortunately, CMOS detectors do not bleed but the stars will be saturated. In this section, we describe a simulation study to answer how we can accurately centroid a saturated star using the diffraction spikes. The parameters used for simulation are displayed in Table 2.

Table 2. Parameters used for simulation study to perform astrometry using diffraction spikes.

| Telescope Diameter (m) | 6 |
| Secondary (m) | 1.8 |
| Width of spider (m) | 0.12 |

From the physical optics point of view, the diffraction spikes are a part of the PSF of the telescope caused by the spider holding up the secondary. The spider blocks light at the pupil of the telescope, thus changing the amplitude of the wavefront. The vertical spider causes the horizontal diffraction spike and vis-versa. The PSF of the telescope, which includes the Airy pattern and diffraction spikes, is the square of the Fourier transform of the electric field E(x,y) = Ap(x,y) exp(i $\phi(x,y)$)), where Ap(x,y) is a circular telescope aperture function with a central obscuration and $\phi(x,y)$ is the wavefront the at the pupil (Goodman 1968). An amplitude perturbation of the E-field of a perfect wavefront produces a symmetric change in the PSF. That is if we move the position of the vertical spider in the pupil, the horizontal diffraction spike (caused by the vertical spider) does not move. What causes the diffraction spike to be not centered are phase errors in the pupil.

3.3.1    Core-spike offset

Wavefront errors in the pupil also affect the centroid of the central lob of PSF relative to the spikes because the phase errors that affect the central lobe are low spatial frequency phase errors like coma while the phase errors that affect diffraction spike are high spatial frequency errors. As a result, we can expect some amount of offset bias between centroiding the stellar image and the diffraction spikes. We shall call this offset between the PSF core centroid and the diffraction spike centroid *core-spike offset*. We did a simulation assuming a random wavefront error with λ/18 P-V error and λ/100 RMS (root-mean-square) with a $1/f^3$ power law shown in Figure 9(a) to quantify the core-spike offset.

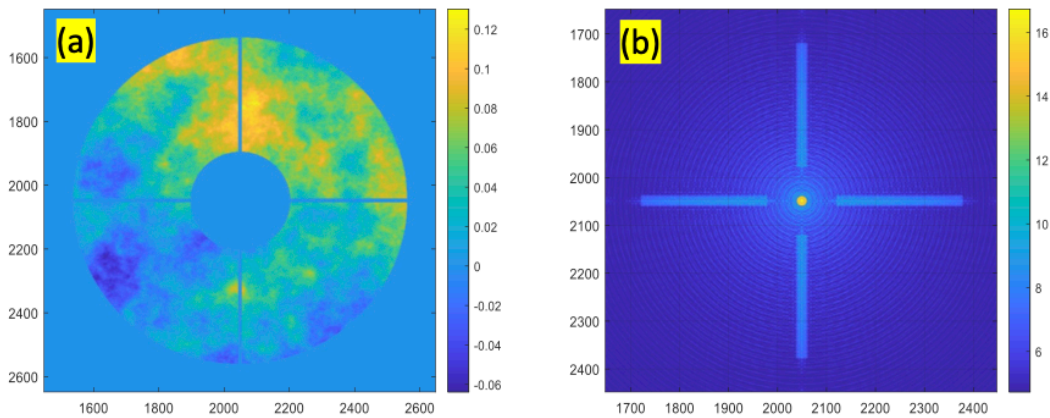

Figure 9. (a) Wavefront errors (radian) over the pupil with an amplitude of l/100 in RMS or  /18 peak-to-valley; (b) point-spread-function in log scale and mask used for centroiding spikes.

The spider covers about 2% of the area of the clear aperture. Because the RMS of the wavefront error is small, the PSF is visually the same as the Airy spot from a perfect telescope. At 0.6um, the Airy spot has an FWHM (full width at the half maximum) of 20 mas for a 6 m telescope. We centroid the image in two different ways, namely the *core centroid*, defined as the centroid of the part of the PSF consisting of the central lob together with the 1st and 2nd diffraction rings, and the *diffraction-spike centroid*, defined as the centroid of diffraction spikes. Figure 9(b) shows the masks used for the core centroid and the diffraction spike centroid. The core mask uses the light in the central lobe as well as the 1st and 2nd diffraction rings.  The diffraction spike mask only used the light from the diffraction spike when the light from the spike is brighter than the 4th circular diffraction ring.

The centroids were calculated by fitting the PSF from a telescope with zero wavefront error to the core and then to the diffraction spikes. Because the wavefront error has some tilt, the core centroid is biased by the tilt and low-order odd aberrations like coma. The diffraction spikes are biased by the average wavefront tilt and high-order wavefront errors. For the case of the wavefront shown in Fig 9a, the biases in uas are listed in Table 3.

Table 3. Centroid offsets due to wavefront aberrations

|  | X ($\mu$as) | Y($\mu$as) | R ($\mu$as) |
| --- | --- | --- | --- |
| Core centroid | 787.34 | -407.69 | 884.64 |
| Diff spike | 663.38 | -180.96 | 687.61 |
| Core-spike offset | 123.97 | -226.74 | 258.42 |

The core-spike offset is a fraction of a milliarcsecond, which is consistent with Gaia's performance of ~100 uas for stars brighter than G=6 mag using the diffraction spikes (Sahlmann et al. 2016). Now if the wavefront error is constant, the core-spike offset would also be constant. In narrow-angle astrometry, we can calibrate this astrometric offset between the central lobe and the diffraction spikes.

### 3.3.2 Calibration of core-spike astrometric bias

Calibration of core-spike astrometric bias can be done by using an appropriately short exposures, such that the PSF peak of the bright star is below saturation for simultaneously centroiding the core PSF and the diffraction spikes to estimate the core-spike offset. For a spider that blocks 2% of the primary area, the surface brightness of the diffraction spike is ~10 magnitudes fainter than the peak based on simulation. From the photon noise point of view, the diffraction spikes are sufficiently bright. For calibration, the diffraction spike is still at ~6e well above the low read noise (~1.5e) of a typical modern CMOS sensor assuming the core PSF is slightly below the saturation at (~60,000e/pixel). For science measurements, the typical reference star brightness is in the range of ~12-18 mag, fainter than the spikes of a saturated star in general, therefore we will not be limited by the photon noise in diffraction-spike centroiding.

We now estimate the time required to calibrate a core-spike offset. Considering the error due to photon noise, centroiding the central lobe of a star image from a 6 m telescope to 1 uas accuracy requires a total of N = 1e8 photons according to the accuracy formula $\lambda/(2*D*sqrt(N))$. With Nyquist sampling, the PSF spread over a bit more than 2x2 pixels, so each image would collect ~200,000 photons. The central core can be centroided to 1uas using ~500 images to get the required 1e8 photons. To centroid the diffraction spikes, we would need 50X as many images, 25,000 images. These stars are extremely bright, and the typical exposure would be 10s of microseconds. When a CMOS sensor only reads out a 256x256 pixel region, the readout can be fast up to 1 KHz, so 25,000 images would take about 25 seconds.

However, the wavefront error will change if we place the target star in different parts of the FOV because of the beam walk on the tertiary mirror of a TMA telescope (Fig. 5). We therefore need to calibrate the core-spike offset accounting for field dependency. Fortunately, it is possible to put target in general near the center of the field, which makes the calibration of the field dependency much easier. The idea is to calibrate the core-spike offsets by putting targets in a grid of locations near the center of the field. As described above, the amount of telescope time needed to calibrate the core-spike offset is pretty short, which would be feasible for real operation.

Our simulation assumes that the telescope has an ~8 arcmin FOV and the target star can be placed at the center of the FOV to +/- 3.4 arcsec, so the beam walk is about +/- (0.5 1/(8x60)*3.4 =) 1/280 of the diameter of the tertiary. Using a similar simulation as described in subsection 3.2.2, where the wavefront error was generated over a 4Kx4K array with a circle of diameter of 1K representing the footprint of the target star. The footprint on the tertiary mirror depends on the target position in the field. We simulate the cases where the target star is put on a 5x5 grid at the center of the field with a grid spacing of 1.7 arcsec (~ 330 pixels for sampling the focal plane of a 6m telescope with a pixel scale of $\lambda/(4D)$). The effect due to beam-walk is simulated by shifting the footprint (circular phase screen) an amount reflected by the beam-walk on the tertiary. We then calculate the centroids of the core PSF and the diffraction spikes using the corresponding wavefront errors. The spike-core offsets for 5x5 points in the FOV are displayed in Fig. 11 respectively for X and Y directions as two colormaps.

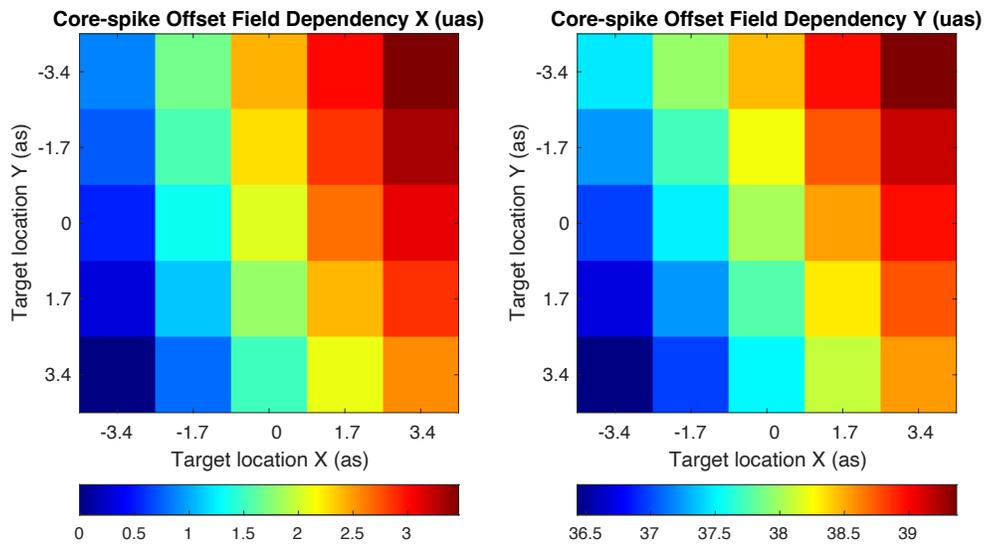

Figure 10. Core-spike offset as function of target location in the field of view.

There is an overall offset of about 38 uas along Y. The variation of the core-spike offset with the target position in the field shows a dominant linear gradient with a range of about 3 uas over the 5x5 grid. A quadratic form $C_0 + C_1 X + C_2 Y + C_3 X^2 + C_4 XY + C_5 Y^2$ can model this dependency with residuals shown in Fig. 12. To the accuracy of 0.1 uas, these residuals are negligible. Therefore, if we estimate $C_i, i = 0,1,\cdots,5$, for both the core-spike offset along X and Y directions respectively using the calibration data by putting the target star at a 5x5 grid point near the center of the field, we can correct the field-dependent core-spike offset for astrometry of the target star using its diffraction spikes.

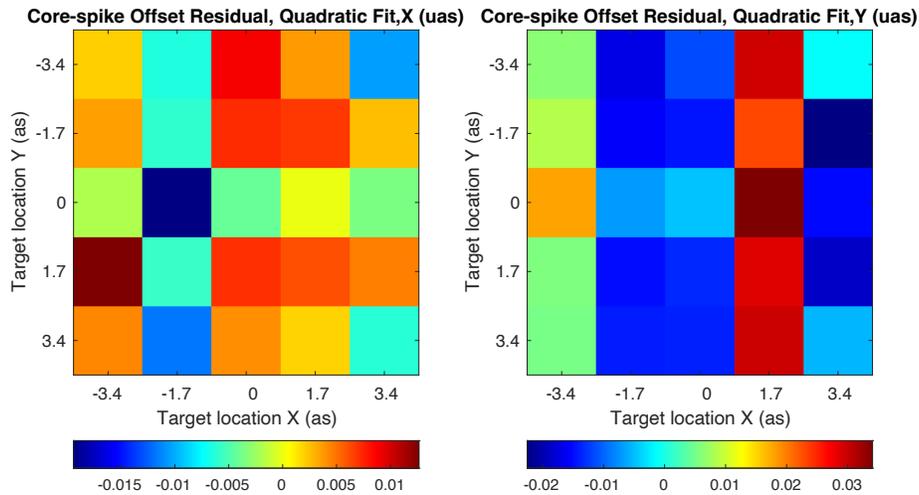

Figure 11. Core-spike offset residuals after fitting a quadratic function of target field position.

## 4 Future work

We have presented an architecture for calibrating systematic errors needed to achieve sub-uas astrometry for a 6 m telescope and demonstrated the key concepts using simulations and some lab experiments. Future work will extend our detector calibration to large format CMOS imaging arrays for a large FOV. We also plan to perform an experimental demonstration of calibrating optical distortion at the 1e-4-pixel level. And last, we would like to do an experimental demonstration of centroiding using diffraction spikes to 1e-4 pixels.

### 4.1 Calibrate pixel responses of a large format CMOS detector

We plan to calibrate the pixel responses of a large-format CMOS detector by extending our existing work on the calibration of an E2V CCD. This extension would require accounting for the fringe curvatures due to the large detector, for which the linear approximation of the fringes is no longer valid.

### 4.2 Test field distortion in lab

We have used simulations to study the field distortion calibration and would like to further validate the concept with lab experiments. The distortion map of the flight telescope can be measured in lab with a holographic element illuminated with a laser light to generate 10,000 points in the focal plane. In the lab demonstration, we will use a Cassegrain telescope instead of a TMA to reduce cost. We will scale the FOV of the system so that the fractional beam walk on the secondary is roughly the same as on the strawman design of the large TMA telescope.

The experimental setup will consist of a laser source and ~ 20cm collimating telescopes as displayed in Fig. 12. The collimated beam will hit a diffracting surface, which is a rectangular array of 60um holes in a nickel coating. The diffracting optic was created at JPL's microdevices lab using their e-beam lithography machine, precise at the few 10's nm level. The diffractive surface produces an array of beams, which the receiving telescope focuses into a rectangular array of Airy disks. Wavefront errors in the collimating telescope will be common across all the images. Any irregularities in the spacing of the holes will cause a slight change in the PSF of the Airy spots but again be common across all the images in the focal plane. The receiving telescope will be a ~20cm Cassegrain telescope on a mount that can tip and tilt the whole telescope and camera.

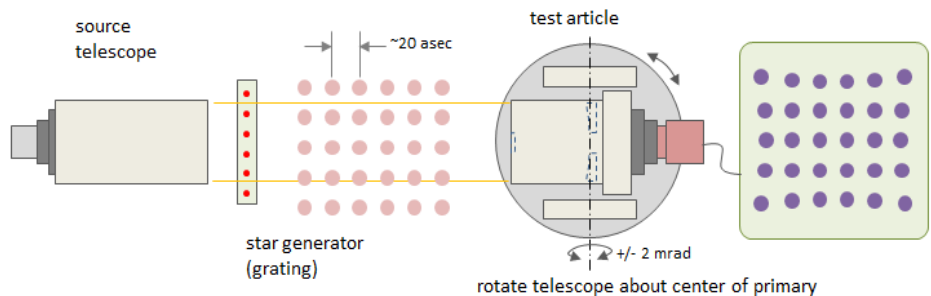

Figure 12. Testbed setup for field distortion calibration. The grating generates a regular pattern (grid) of stars. The test article represents the flight telescope being calibrated for field distortion. A pincushion distortion is illustrated.

The diffractive grating is designed to produce a brighter zeroth-order diffractive image than the others. The zeroth-order diffractive image will be the "target" star, and the others will serve as reference stars. Because the reference stars are generated by the diffraction grating, we know their relative positions precisely. As a result, we can use their centroids to compute the coefficients of the polynomial distortion model directly, which we expect to compensate for both the changes in telescope alignment and the $\lambda/20$ P-V figure errors of the optics. Centroiding to 1e-4 pixels requires getting enough signal so that photon noise level is below 1e-4, therefore, we need to collect at least 1e8 photons on the target star. Since the full well is ~50,000e, and the PSF spans ~ 2*2 pixels, we need ~500 exposures. If we record images at ~1Hz, this will take ~500 sec per data set.

We will first take one data set, solve for the polynomial coefficients, and measure the position of the target star. We will then tip/tilt the telescope/camera by a few arcmins and collect another data set (~500 images). We will repeat the analysis to verify that the target star is at the same location relative to the reference stars. We will do this experiment in a vacuum chamber to validate astrometric accuracy at the 1e-4-pixel level by repeating the tip/tilt 10 times and checking the consistency of the estimated target star's position relative to reference stars. A typical atmospheric seeing causes a fraction of arcsec motion for stellar images from a ground-based telescope. Averaging 500 images might bring the atmospheric image motion to ~ 10 mas. In a lab setup shielded from the heating/air-conditioning airflow, the air turbulence is at least 10 times smaller. In a vacuum chamber, we expect the atmospheric turbulence effect to be less than 1uas at a pressure below 1 mbar. The vacuum chambers we plan to use will use roughing pumps to get pressure below ~10 ubar.

### 4.3 Centroiding using diffraction spikes

We also plan to demonstrate in the lab the centroiding of saturated stars using diffraction spikes. We will project light from a fiber into a simple telescope with a fake spider to create the diffraction spikes. To calibrate the offset between the Airy core centroid and the diffraction spike centroid, we will use a CMOS detector that allows a very rapid readout of a small part of the chip. The IMX455 allows reading out a 300x300 pixel region at 500Hz. For a Nyquist sampled PSF and a ~50,000e full well, we collect ~200,000 photons of the star and more than 2000 photons for the diffraction spike per image. Calibrating the offset to 1e-4 pixels would require collecting 1e8 photons from the diffraction spikes. We will need 50,000 images, which would be 100 sec of data at 500Hz.

As a validation, we will move the image to another part of the detector by translating the detector by a few 10's to 100 pixels away and repeat to verify that we have measured the core-spike offset with adequate accuracy. We will first calibrate the detector's pixel responses and incorporate the effective pixel locations in the data processing to generate centroid estimations (Zhai et al. 2011).

## 5 Conclusions

To achieve uas astrometric measurements with space telescopes, we need to account for the effects of everything that photons touch traveling from the target star to the detector. In this paper, we have outlined an approach for uas-level narrow-angle relative astrometry and presented three technologies to calibrate errors due to detector, optics, and star saturation. The detector errors due to pixel geometry and QE gradients within a pixel can be calibrated with laser metrology. Optical errors lead to field distortion errors that can be modeled as low-order 2D polynomials and calibrated by observing dense star fields with dithers. Star saturation errors can be described by the "core-spike" offset, the astrometric offset between the centroid of a star's core and the centroid of the diffraction spikes, which can be calibrated using high frame rate images.

We have presented lab results showing 1e-5 $\lambda$/D centroiding accuracy by using laser metrology calibration and discussed the extension to larger detectors. We have also presented field distortion results from modeling high-quality optics with wavefront errors of a typical $1/f^3$ power spectrum and the calibration of the errors observing 100 reference stars. We also discussed how we could demonstrate field distortion calibration in the lab. And last of all, for astrometry of bright target stars relative to faint reference stars, we can calibrate the core-spike offset and its weak fiend dependency. This core-spike offset calibration allows us to use longer exposures to measure the position of bright stars relative to faint reference stars to avoid excessive read noise by centroiding the diffraction spikes of the saturated stars. We hope these technologies will mature for future missions like the 6 m flagship mission to adopt and enable the new science capabilities from the uas astrometry.

Acknowledgements